# Heavy-Meson Masses in the Framework of Trigonometric Rosen-Morse Potential Using the Generalized Fractional Derivative

M. Abu-Shady[1] and Etido P. Inyang[2]

[1] Department of Mathematics and Computer Science, Faculty of Science, Menoufia University, Egypt

Email: dr.abushaddy@gmail.com

[2] Department of Physics, National Open University of Nigeria, Jabi, Abuja, Nigeria
Email: einyang@noun.edu.ng

## Abstract

*Trigonometric Rosen-Morse Potential is employed as a mesonic potential interaction. The extended Nikiforov-Uvarov method is used to solve the N-radial Fractional Schrödinger equation analytically. Using the generalized fractional derivative, the energy eigenvalues are obtained in the fractional forms. The current findings are used to calculate the masses of mesons such as charmonium, bottomonium, and heavy-light mesons. The current findings are superior to those of other recent studies and show good agreement with experimental data as a result, the fractional parameter is crucial in optimising meson masses.*

## Introduction

Hadrons are made up of quarks and gluons, which are more fundamental particles. Quantum chromodynamics (QCD) is the quantum theory of these particles. This theory has vital properties such as asymptotic freedom, spontaneous symmetry breaking, and confinement. Because of the asymptotic freedom, perturbative QCD calculations work in high-energy physics. The low-energy effective field theory is used to examine non-perturbative QCD phenomena. The properties of the heavy and light hadrons can be studied in two ways: the relativistic quark models [1-4] and the non-relativistic quark models [5-10]. The Schrödinger equation (SE) is used as a non-relativistic quark model. As a result, the SE solutions are critical for computing quarkonium masses. For specific potentials, numerous methods have been utilised to find exact and approximate SE solutions, such as the Nikiforov–Uvarov (NU) method [9-13] among others [14-16]. The trigonometric Rosen-Morse potential (TRMP) has recently gained popularity as a useful potential for QCD. The potential is used in one-dimensional space according to references [17, 18]. Deta et al. [19] used the NU approach to expand the investigation to D-dimensional space. According to Abu-Shady and Ezz-Alarb [6], the trigonometric quark confinement potential is an effective tool for quark model estimates of heavy meson properties using exact-analytical iteration method.

The fractional calculus has recently garnered attention in various domains of physics that feature nonlinear, complex phenomena, such as Refs. [11, 12, 20] where the authors used the conformable fractional derivative to solved the fractional N-dimensional radial SE with extended Cornell potential by using the extended Nikiforov-Uvarov (ENU) method. Recently, Abu-shady and Kaabar [21] suggested a new definition for fractional derivative called the generalized fractional derivative and it coincides with classical fractional derivative and also a simple tool for fractional differential equations

This work aim to employ the TRMP to calculate the mass spectra(MS) of heavy and heavy-light meson in the framework of fractional SE by using the generalized fractional extended Nikiforov–Uvarov method (GFD-ENU).

**2-Trigonometric Rosen- Morse potential**

The features of TRMP is discuss in this section, and it takes the form [6, 22].

$$V(z) = \frac{1}{2\mu d^2}\left(-2b\cot|z| + \frac{a(a+1)}{(Sin|z|)^2}\right), \tag{1}$$

where $a$ = 1, 2, 3, and $z = r/d$. $\mu, b, d$ are parameters to be determined later. We expand the potential in a Taylor series for small z and have,

$$V(r) = -\frac{A}{r} + Br + \frac{C}{r^2} + Dr^2, \tag{2}$$

where,

$$A = \frac{b}{\mu d}, B = \frac{b}{3\mu d^3}, C = \frac{a(a+1)}{2\mu}, \text{ and } D = \frac{a(a+1)}{30\mu d^4} \tag{3}$$

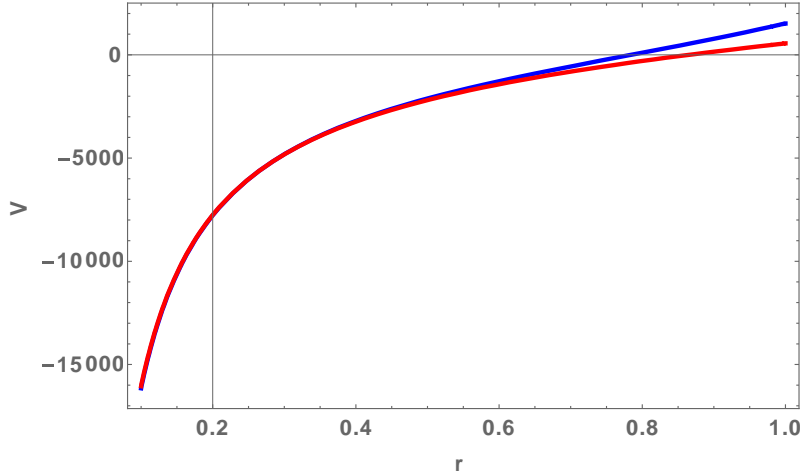

**Fig. (1):** Plot of TRMP for the exact and approximate potential.

The plot of the TRMP as a function of r is shown in Fig.1, we note that TRMP has two features the Coulomb potential (CP) and confinement potential, the short distance is describe by CP and the long distance is describe by the confinement force. The approximate and exact potentials concise at point up to 0.8 fm. Therefore, TRMP is a good potential for the depiction of quarkonium interaction.

**3-Generalized Fractional Derivative of the Schrödinger Equation**
The SE in the *N*-dimensional space is of the form [7].

$$\left[\frac{d^2}{dr^2} + \frac{N-1}{r}\frac{d}{dr} - \frac{L(L+N-2)}{r^2} + 2\mu(E_{nl} - V(r))\right]\psi(r) = 0, \tag{4}$$

where L, *N*, and μ are the angular momentum quantum number, the dimensionality number, and the reduced mass, respectively. We set the wave function $\psi(r) = r^{\frac{1-N}{2}} R(r)$, and Eq. (4) takes the form

$$\left[\frac{d^2}{dr^2} + 2\mu(E_{nl} - V(r)) - \frac{\left(L+\frac{N-2}{2}\right)^2 - \frac{1}{4}}{r^2}\right]R(r) = 0, \tag{5}$$

By substituting Eq. (2) into Eq. (5), we obtain the following equation.

$$\left[\frac{d^2}{dr^2} + 2\mu\left(E_{nl} + \frac{A}{r} - Br - \frac{C}{r^2} - Dr^2\right) - \frac{\left(L + \frac{N-2}{2}\right)^2 - \frac{1}{4}}{r^2}\right]R(r) = 0, \tag{6}$$

Eq. (6) is compact in the form

$$\left[\frac{d^2}{dr^2} + \frac{1}{r^2}\left(\varepsilon r^2 + A_1 r - B_1 r^3 - C_1 - D_1 r^4\right)\right]R(r) = 0, \tag{7}$$

where,

$$\varepsilon = 2\mu E_{nl},\ A_1 = 2\mu A,\ B_1 = 2\mu B, \tag{8}$$

$$C_1 = 2\mu C + \left(L + \frac{N-2}{2}\right)^2 - \frac{1}{4},\ D_1 = 2\mu D. \tag{9}$$

To put Eq. (7) in the fractional form, firstly, we introduce the dimensionless form by introducing $Z = \frac{r}{w}$ where $w$ is scaler quantity measure in GeV

$$\left[\frac{d^2}{dz^2} + \frac{1}{z^2}\left(\varepsilon z^2 + A_1 z - B_1 z^3 - C_1 - D_1 z^4\right)\right]R(z) = 0, \tag{10}$$

where

$$\left.\begin{array}{l}\varepsilon = \dfrac{2\mu E_{nl}}{w^2},\ A_1 = \dfrac{2\mu A}{w},\ B_1 = \dfrac{2\mu B}{w^3}, \\ C_1 = 2\mu C + \left(L + \dfrac{N-2}{2}\right)^2 - \dfrac{1}{4},\ D_1 = \dfrac{2\mu D}{w^4}\end{array}\right\} \tag{11}$$

We write generalized fractional derivative Eq. (10) as in [12],

$$\left[D^\alpha D^\alpha + \frac{1}{z^{2\alpha}}\left(\varepsilon z^{2\alpha} + A_1 z^\alpha - B_1 z^{3\alpha} - C_1 - D_1 z^{4\alpha}\right)\right]R(z) = 0, \tag{12}$$

where the potential defines in Eq. (2) is replaced by

$$V(z) = -\frac{A}{z^\alpha} + Bz^\alpha + \frac{C}{z^{2\alpha}} + Dz^{2\alpha}, \tag{13}$$

by using the generalized fractional derivative in Ref. [21] and also Ref. [12].

$$D^\alpha\left[D^\alpha R(z)\right] = \lambda^2\left[(1-\alpha)z^{1-2\alpha}R'(z) + s^{2-2\alpha}R''(z)\right], \tag{14}$$

where

$$\lambda = \frac{\Gamma(\beta)}{\Gamma(\beta-\alpha+1)} \text{ with } 0 \prec \alpha \leq 1 \text{ and with } 0 \prec \beta \leq 1 \tag{15}$$

By substituting Eq. (14) into (12), we get

$$R''_{nl}(z) + \frac{1-\alpha}{z} R'_{nl}(z) + \frac{1}{z^2}\left(\varepsilon z^{2\alpha} + A_1 z^\alpha - B z^{3\alpha} - C_1 - D_1 z^{4\alpha}\right) R(z) = 0, \tag{16}$$

where

$$\left.\begin{array}{l} \bar{\tau}_f(s) = 1-\alpha, \sigma(s) = z, \\ \tilde{\sigma}(s) = \frac{1}{\lambda^2}(\varepsilon_{nl} Z^{2\alpha} + A_1 Z^\alpha - B_1 Z^{3\alpha} - C_1 - D_1 Z^{4\alpha}), \\ \lambda = \frac{\Gamma(\beta)}{\Gamma(\beta-\alpha+1)} \end{array}\right\} \tag{17}$$

By using ENU-CFD method [11], we obtain $\pi_f$

$$\pi_f = \frac{\alpha}{2} \pm \sqrt{\frac{\alpha^2}{4} - \frac{1}{\lambda^2}(\varepsilon_{nl} Z^{2\alpha} + A_1 Z^\alpha - B_1 Z^{3\alpha} - C_1 - D_1 Z^{4\alpha}) + zG(z)}. \tag{18}$$

To choose the function form of $G(z)$. We then find a function $G(z) = S + Pz^{\alpha-1} + Qz^{2\alpha-1}$ that makes the function under the root in the above equation to become quadratic

$$\pi_f = \frac{\alpha}{2} \pm \sqrt{(A_{11} + B_{11} Z^\alpha + F_{11} Z^{2\alpha})^2}. \tag{19}$$

then

$$\pi_f = \frac{\alpha}{2} \pm \left(A_{11} + B_{11} Z^\alpha + F_{11} Z^{2\alpha}\right). \tag{20}$$

Substituting $G(z)$ in Eq. (18) and then equating the coefficients in Eqs. (18 and 20). To find the constants $S, P, Q, A_{11}, B_{11},$ and $F_{11}$

$$(A_{11} + B_{11} Z^\alpha + F_{11} Z^{2\alpha})^2 = \frac{\alpha^2}{4} - \frac{1}{\lambda^2}(\varepsilon_{nl} Z^{2\alpha} + A_1 Z^\alpha - B_1 Z^{3\alpha} - C_1 - D_1 Z^{4\alpha})$$
$$+ z\left(s + pz^{\alpha-1} + Qz^{2\alpha-1}\right) \tag{21}$$

so, we obtain

$$(A_{11}^2 - \frac{\alpha^2}{4} - \frac{C_1}{\Gamma^2}) + (2A_{11}B_{11} + A_1 - p)Z^\alpha + (B_{11}^2 + 2A_{11}F_{11} + \varepsilon_{nl} - Q)Z^{2\alpha} + (2B_{11}F_{11} - B_1)Z^{3\alpha}$$
$$+(F_{11}^2 - D_1)Z^{4\alpha} = 0 \tag{22}$$

then, we have

$$A_{11} = -\frac{1}{2}\sqrt{\alpha^2 + \frac{4c_1}{\lambda^2}}, \tag{23}$$

$$B_{11} = \frac{-A_1 + p}{2A_{11}} = \frac{-A_1 + p}{\sqrt{\alpha^2 + \frac{4c_1}{\lambda^2}}}, \tag{24}$$

$$F_{11}^2 - D_1 = 0 \Rightarrow F_{11} = \sqrt{D_1}, \tag{25}$$

$$2B_{11}F_{11} - B_1 = 0 \Rightarrow B_{11} = \frac{B_1}{2\sqrt{D_1}} \tag{26}$$

$$B_{11}^2 + 2A_{11}F_{11} + \varepsilon_{nl} - Q = 0 \Rightarrow \varepsilon_{nl} = Q - B_{11}^2 - 2A_{11}F_{11} \tag{27}$$

By using Eqs. (23–26), we have

$$\varepsilon_{nl} = -\frac{B_1^2}{4D_1} + \sqrt{D_1}\left[2\alpha(1+n) + \sqrt{\alpha^2 + \frac{4}{\lambda^2}\left(2\mu C + \left(l + \frac{N-2}{2}\right)^2 - \frac{1}{4}\right)}\right] \tag{28}$$

from Eq. (28), we obtain the energy spectrum of the TRMP

$$E_{nl} = -\frac{B^2}{4D} + \sqrt{\frac{2D}{\mu}}\left[2\alpha(1+n) + \sqrt{\alpha^2 + \frac{4}{\lambda^2}\left(2\mu C + \left(l + \frac{N-2}{2}\right)^2 - \frac{1}{4}\right)}\right] \tag{29}$$

## 3. Mass spectra of heavy and heavy-light mesons

The MS of the heavy and heavy-light mesons is predicted in the 3-dimensional (N=3), by applying the relation [23, 24].

M=$m_q$+ $m_{\bar{q}}$ +$E_{nl}$ , (30)

where m is bare quark mass for quarkonium.

Plugging Eq. (29), into Eq. (30) gives:

$$M = m_q + m_{\bar{q}} - \frac{B^2}{4D} + \sqrt{\frac{2D}{\mu}} \left[ 2\alpha(1+n) + \sqrt{\alpha^2 + \frac{4}{\lambda^2}\left(2\mu C + \left(l + \frac{N-2}{2}\right)^2 - \frac{1}{4}\right)} \right] \qquad (31)$$

where $\lambda = \frac{\Gamma(\beta)}{\Gamma(\beta-\alpha+1)}, 0 \prec \alpha \leq 1$ and $0 \prec \beta \leq 1$

### 4-Results and Discussion

Using Eq. (31), we determined the MS of charmonium for the states 1S to 4S as displayed in Table (1). The experimental data (ED) are used to fit the free parameters of the current computations, B, D, and C. Additionally, bare quark masses may be found in Ref. [6]. By calculating total error in comparison with ED, we observe that estimations of charmonium masses are in good agreement with ED and are improved over Refs. [6, 25, 26] that used different potentials and methods. Table (2) show that the bottomonium spectrum masses for states 1S to 4S coincide with ED and that the current calculations are more accurate than those in Refs. [6, 25-27], with a smaller total error.

For states from 1S to 3S, we use the $2m = m_b + m_c$ part of Eq. (31) to calculate the MS of heavy-light ($b\bar{c}$) mesons. We notice that the 1S and 2S states are closer, but the ED for the other states are not known. By calculating the sum error for these works, we can see that the current predictions of the $b\bar{c}$ mass are better than those in Refs. [6, 28, 29]. Using the $2m = m_c + m_s$ part of Eq. (31), we compute the MS of the $c\bar{s}$ mesons from the 1S to the 1D state in Table (4). Close to ED are 1S and 2S. By computing the total error for each potential, other states are enhanced in contrast to the screened potential and phenomenological potential [30].

In the current work, we observe that for every state, as shown in Tables (1, 2, 3, 4), the MS increase as N is increased at N = 4 for various states. This resulted in the restriction of non-relativistic quark models for higher dimensional. The plots of the eigenvalues against the principal quantum number ($n$) are shown in Figures (2-5). In Fig. (2), it was noticed that as the dimensional number increases, the curves of the energy increase at different values of $n$. In addition, the curves of energy are shifted to higher values by increasing N. The variation of eigenvalues of $c\bar{c}$ against $n$ for different values of $\alpha = 0.2, \alpha = 0.5,$ and $\alpha = 0.7$ is shown in Fig. (3). We note the energy increases when $n$ is increased. Also, we note that an increase in the fractional parameter leads to decreasing energy eigenvalue for $b\bar{c}$ meson and $c\bar{s}$ meson as in Figs. (4, 5). Thus, the fractional parameter make the energy more stable.

**Table (1)**. Mass spectra of charmonium (in GeV), ( $m_c$=1.275 GeV [31], µ =0.638 GeV, $D = 0.2630$ GeV$^{-1}$, $B = 1.4925$ GeV$^2$, $N = 3, \alpha = 0.1$ $\beta = 1$, $C = 4.7598$

| State | Present paper | [25] | [26] | [27] | [6] | N=4 | Exp. [31] |
|---|---|---|---|---|---|---|---|
| 1S | 3.096 | 3.078 | 3.096 | 3.078 | 3.239 | 3.352 | 3.096 |
| 1P | 3.421 | 3.415 | 3.433 | 3.415 | 3.372 | 3.533 | 3.525 |
| 2S | 3.649 | 4.187 | 3.686 | 3.581 | 3.646 | 3.675 | 3.649 |
| 1D | 3.757 | 3.752 | 3.767 | 3.749 | 3.604 | 4.158 | 3.769 |
| 2P | 3.899 | 4.143 | 3.910 | 3.917 | 3.779 | 3.715 | 3.900 |
| 3S | 4.040 | 5.297 | 3.984 | 4.085 | 4.052 | 4.684 | 4.040 |
| 4S | 4.399 | 6.407 | 4.150 | 4.589 | 4.459 | 5.463 | 4.415 |
| Total Error | 0.00521 | 0.9513 | 0.11065 | 0.11152 | 0.05788 | 0.09147 | |

**Table (2).** Mass spectra of bottomonium (in GeV), ( $m_b$= 4.18 GeV [31], $\mu$ =2.09 GeV, $D = 0.347 \text{ GeV}^{-1}$, B = 0.5231 GeV², $\alpha = 0.1$, $C = -0.0154$, $\beta = 1$

| State | Present paper | [25] | [27] | [26] | [6] | N=4 | Exp. [31] |
|---|---|---|---|---|---|---|---|
| 1S | 9.445 | 9.510 | 9.510 | 9.460 | 9.495 | 9.552 | 9.444 |
| 1P | 9.915 | 9.862 | 9.862 | 9.840 | 9.657 | 10.458 | 9.900 |
| 2S | 10.023 | 10.627 | 10.038 | 10.023 | 10.023 | 10.567 | 10.023 |
| 1D | 10.162 | 10.214 | 10.214 | 10.140 | 10.161 | 10.453 | 10.161 |
| 2P | 10.246 | 10.944 | 10.390 | 10.160 | 10.26 | 10.564 | 10.260 |
| 3S | 10.355 | 11.726 | 10.566 | 10.280 | 10.355 | 10.473 | 10.355 |
| 4S | 10.553 | 12.834 | 11.094 | 10.420 | 10.579 | 10.654 | 10.579 |
| Total Error | 0.000917 | 0.488529 | 0.09777 | 0.08207 | 0.09612 | 0.03086 | |

**Table (3).** Mass spectra of b$\bar{c}$ meson (in GeV) ($m_b$=4.18 GeV, $m_c$=1.275 GeV, $\mu$ = 0.976 GeV, D = 0.23, B = 2.1150 GeV², $\alpha = 0.1$, $\beta = 1$ C = 8.0028)

| State | Present paper | [28] | [29] | [6] | N=4 | Exp.[32] |
|---|---|---|---|---|---|---|
| 1S | 6.274 | 6.349 | 6.264 | 6.268 | 6.543 | 6.274 |
| 1P | 6.711 | 6.715 | 6.700 | 6.529 | 6.680 | - |
| 2S | 6.872 | 6.821 | 6.856 | 6.895 | 7.250 | 6.871 |
| 2P | 7.103 | 7.102 | 7.108 | 7.156 | 6.818 | - |
| 3S | 7.179 | 7.175 | 7.244 | 7.522 | 8.127 | - |
| Total Error | 0.00002 | 0.01922 | 0.00377 | 0.00444 | 0.0490 | |

**In Table** (4). Mass spectra of c$\bar{s}$ meson in (GeV) (($m_c$=1.275, $m_s$=0.419) GeV, $\mu = 0.315$ GeV,, $\alpha = 0.1$ , $B = 5.319 \ GeV^2$, $D = 0.23 \ GeV^{-1}$ $\beta = 1$, $C = 275.338$)

| State | Present paper | [6] | Screened | Phenomenological | N=4 | Exp.[32] |
|---|---|---|---|---|---|---|
| 1S | 1.968 | 1.969 | 1.9685 | 1.968 | 2.122 | 1.968 |
| 1P | 2.211 | 2.126 | 2.7485 | 2.566 | 2.364 | 2.112 |
| 2S | 2.318 | 2.318 | 2.8385 | 2.815 | 2.556 | 2.317 |
| 3S | 2.834 | 2.667 | 3.2537 | 3.280 | 3.153 | 2.700 |
| 1D | 2.359 | 2.374 | - | - | 2.797 | 2.318 |
| Total Error | 0.02827 | 0.04392 | 0.73081 | 0.64471 | 0.13502 | |

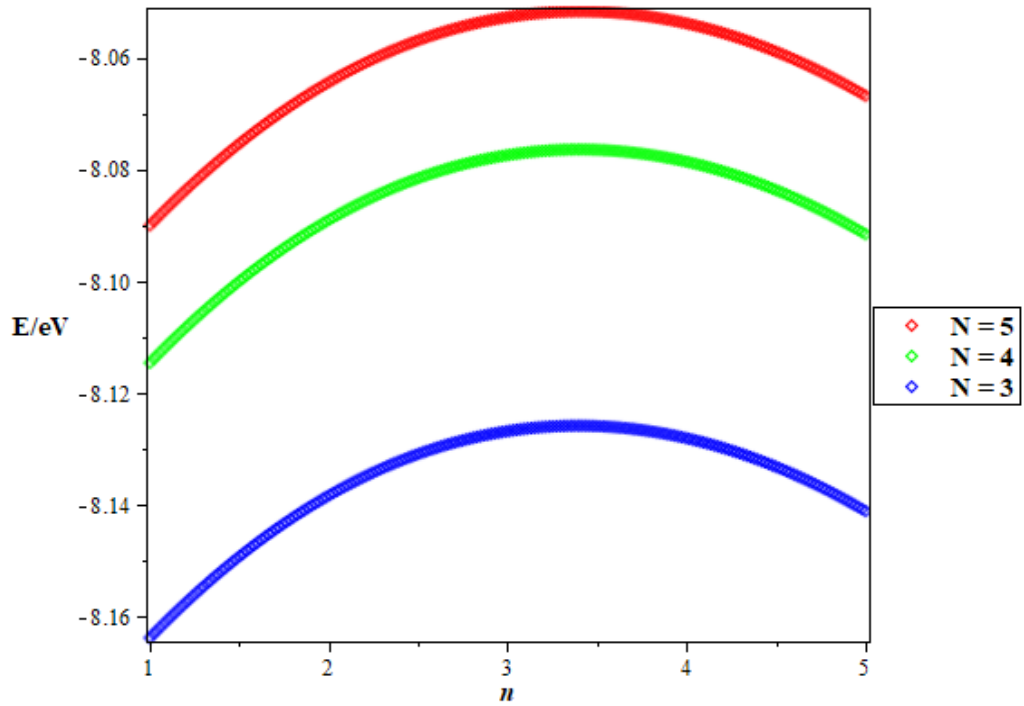

**Fig. (2).** Variation of the energy eigenvalue with $n$ for different dimensional numbers N

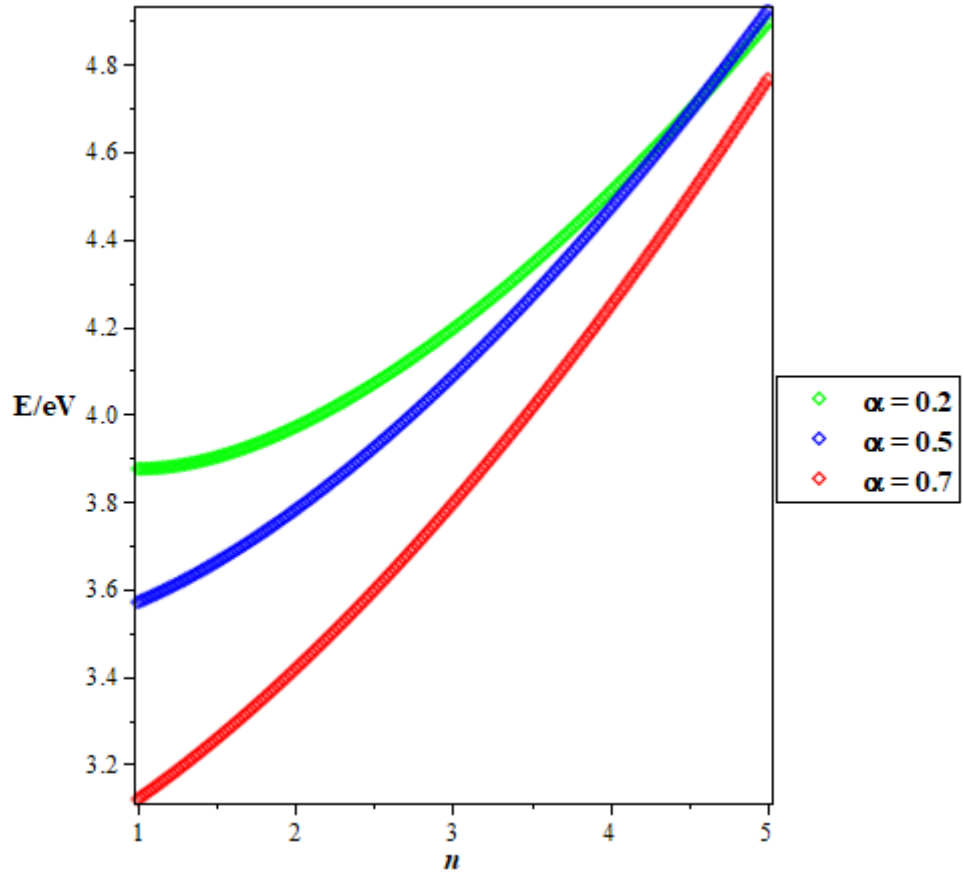

**Fig. (3).** Variation of the energy eigenvalue of charmonium with $n$ for different values of $\alpha$

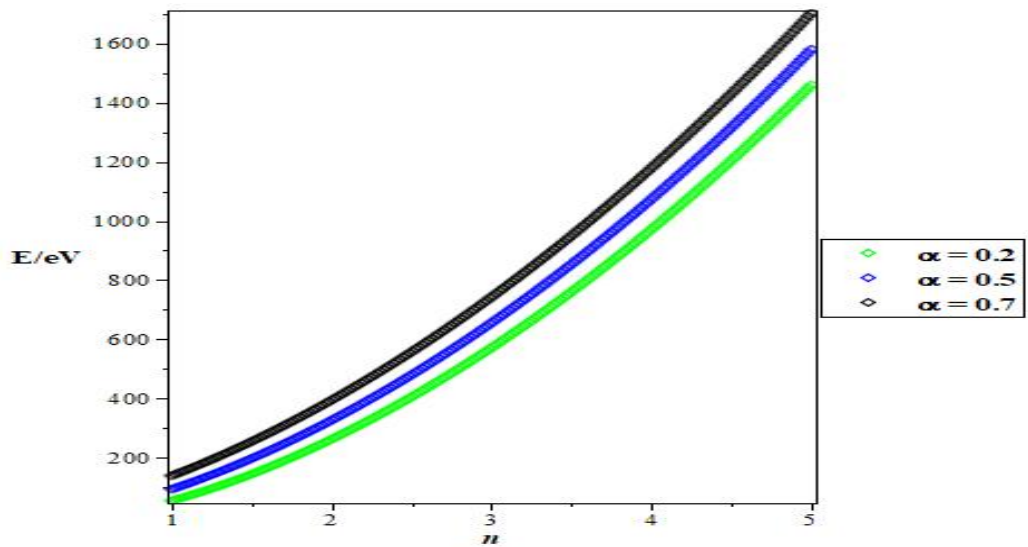

**Fig. (4).** Variation of the energy eigenvalue of $b\bar{c}$ meson with $n$ for different values of $\alpha$

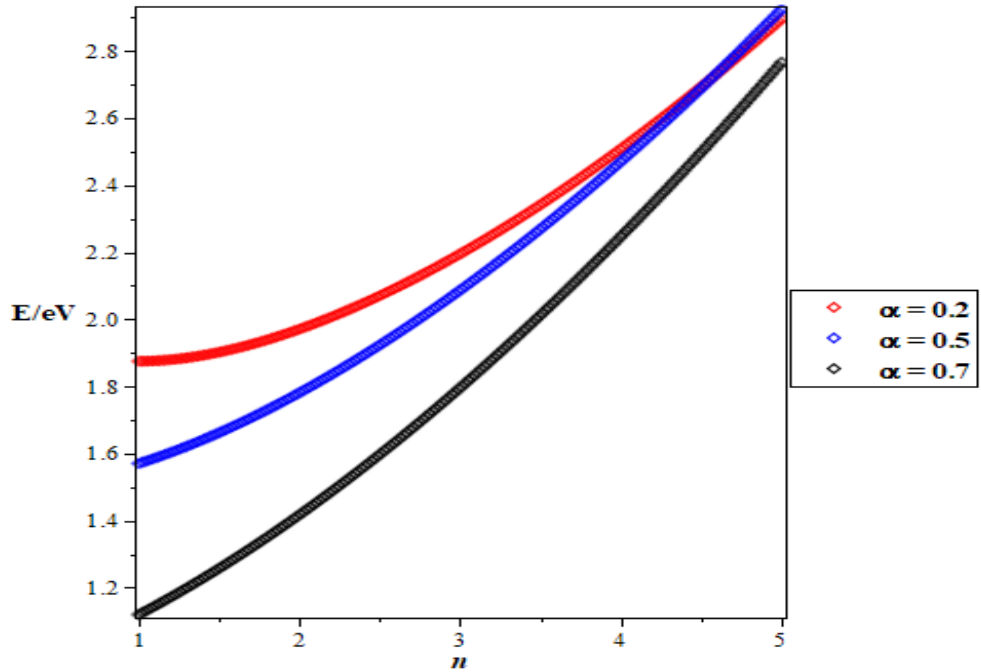

**Fig. (5).** Variation of the energy eigenvalue of $c\bar{s}$ meson with $n$ for different values of $\alpha$

## 4. Conclusion

In this study, the TRMP is model to interact in a quarkonium system and the solutions of SE were obtained using the GFD-ENU method. The energy spectrum was obtained and used to predict the MS of the HMs and heavy-light mesons. The results obtained showed an improvement when compared with the work of other researchers and excellently agreed with ED with a total error less than the errors obtained from other results as shown in Tables 1-4. Also, the figures show that the decreasing fractional parameter leads to lower the energy eigenvalues which makes the state more stable in the fractional model.